# Sub-wavelength focusing meta-lens

**Tapashree Roy, Edward T F Rogers and Nikolay I Zheludev**[*]

Optoelectronics Research Centre & Centre for Photonic Metamaterials, University of Southampton, Southampton, SO17 BJ, United Kingdom

Email: [*]niz@orc.soton.a.c.uk

**Abstract:** We show that planar a plasmonic metamaterial with spatially variable meta-atom parameters can focus transmitted light into sub-wavelength hot-spots located beyond the near-field of the metamaterial. By nano-structuring a gold film we created an array of meta-lenses generating foci of 160 nm (0.2λ) in diameter when illuminated by a wavelength of 800 nm. We attribute the occurrence of sub-wavelength hotspots beyond the near field to the phenomenon of superoscillation.

Metamaterials are known to promise sub-wavelength focusing through negative refraction [1]. Here we focus light into sub-wavelength spots with metamaterials, but without making use of negative refraction. Recently it was shown that precisely tailored diffraction of light on a binary mask can create a sub-wavelength optical hotspot that can be used for optical imaging with resolution far exceeding that of conventional optical instruments [2]. This is possible due to a phenomenon known as superoscillation [3] where interference of propagating components of light may create arbitrarily small hot spots. In practice, however, a superoscillatory binary mask has not been demonstrated to deliver hotspots smaller than λ/3, where λ is the wavelength of light. Superoscillatory masks with the ability to continuously control intensity and phase of the transmitted wave has been reported to create hotspots of any size [4]. Unfortunately manufacturing of such masks, which should have optical thickness and density defined with nanoscale lateral resolution across a device of tens of microns, is still an unattainable technological challenge. Here we show that planar metamaterials can provide a way to manufacture a superoscillatory mask with spatially variable transmission and phase. We illustrate this with a simple example of metamaterial superoscillatory lens manufactured on a plasmonic metamaterial.

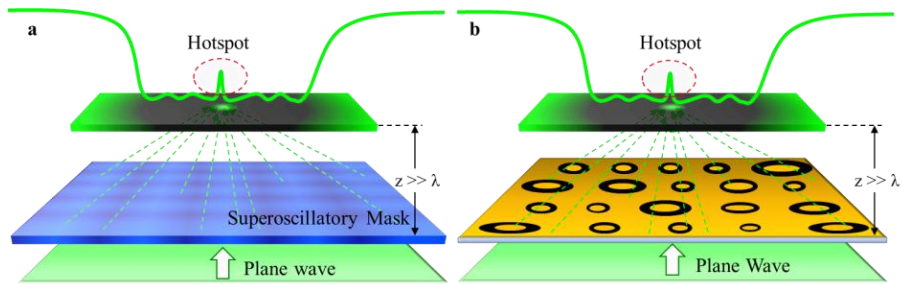

**Figure 1.** Comparison of design and performance of (a) an hypothetical continuous superoscillatory mask, and (b) superoscillatory meta-lens generator, both producing arbitrarily small hot spots at post-evanescent distances from the surface.

Planar metamaterial arrays with spatially varying parameters have attracted attention as wavefront correction devices [5] and wavefront control has been demonstrated in the infrared part of the spectrum [6]. Flat lenses designed with nano-antennas have been demonstrated at telecommunication wavelength to give diffraction limited performance [7]. In this paper we aim towards focal spots smaller than the fundamental diffraction limit at distances considerably beyond the near field of the lens. We choose polarization independent meta-atoms as our design unit to demonstrate optical superoscillation. Other methods for achieving superoscillation, aside from the already mentioned superoscillatory lens [2], and continuous mask [4], includes precisely engineered arrangements of waveguides [8-10]. For these technologies, the focusing devices measure several microns and are difficult to fabricate. In this paper we present a simple design of planar lenses made up of nanometer scale meta-atoms which can be arranged to produce an array of superoscillatory hotspots. Array of lenses made of polystyrene microspheres has already been demonstrated as useful for making repetitive micrometer patterns [11]. We envision our nano-scale meta-lens arrays producing sub-wavelength hotspots will have better uses for such photolithographic processes, and will also find application in high resolution parallel imaging, and optical data storage. In the following sections we discuss the working principle, design, experimental methods and observations for our meta-lens samples.

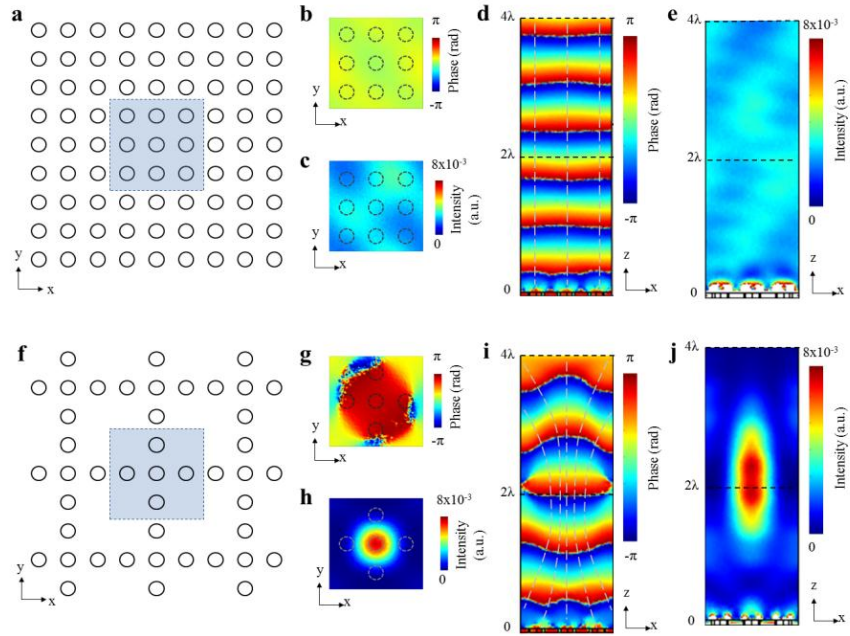

**Figure 2.** Regular metamaterial vs. meta-lens array: (a) Section of an infinite planar array of ring meta-atoms. (b) Phase and (c) intensity, over 3 x 3 meta-atoms, at 2λ from the surface. (d) Phase and (e) intensity in the propagation direction over the 3 x 3 meta-atoms. (f) Section of an infinite meta-lens array. (g) Curved phase front and (h) focussed intensity profile over a meta-lens unit, at 2λ from the surface. Note that phase wrapping occurs in (g). (i) Phase profile like that of a converging lens and (j) intensity showing a focal spot in the propagation direction over a meta-lens unit.

The principle of our metamaterial superoscillatory generator is illustrated on figure 1 in comparison with a conceptual superoscillatory continuous mask [4]. An arbitrarily small hot spot is produced by plane wave incident on a hypothetical superoscillatory mask. The mask is conceptually designed with lateral resolution on the wavelength scale enabling continuous control of the transmission and retardation. In comparison the metamaterial superoscillatory generator exploits the resonant behaviour of the individual sub-wavelength meta-atoms. Each of the meta-atoms can scatter light with defined amplitude and phase depending on its design and position in an array, resulting in transmission with any arbitrary modulation. Indeed the spectral dependencies of phase retardation and intensities are in general different and thus, in principle, it is possible to design a meta-atom with prescribed combination of scattering characteristics. In practice, however, this is difficult and is complicated by three factors namely, absorption losses, finite size of the arrays of meta-atoms and interactions between them. In the view of these challenges we decided to design the metamaterial focusing device using an empirical combinatorial approach [12] and compared the results against fully three-dimensional finite-element simulations (Comsol Multiphysics).

To design the meta-lens we used arrays of meta-atoms which were ring slots in a thin metal film. If an infinite array of meta-atoms is placed with regular sub-wavelength periodicity (figure 2a) and illuminated by a plane wave, the transmitted wave retains a plane wavefront as illustrated on figure 2b-e. For arrays constructed of clusters of five ring slots (figure 2f), light passing through the central part of the cluster experiences a different phase delay than that passing through its outer area: the transmitted light wavefront becomes curved. The wave front curvature is controlled by the ring dimensions, their mutual position in the cluster and depends on the incident wavelength and polarisation. As the central part of the ring cluster provides higher phase delay than its outer part, it forms a convex meta-lens and transmitted light converges to a focus several wavelengths away from the meta-lens surface. A hot-spot is formed directly above the central ring, as illustrated on figure 2h. The planar array of foci repeats itself at fixed distances along the propagation direction, as seen in the classical Talbot effect [13].

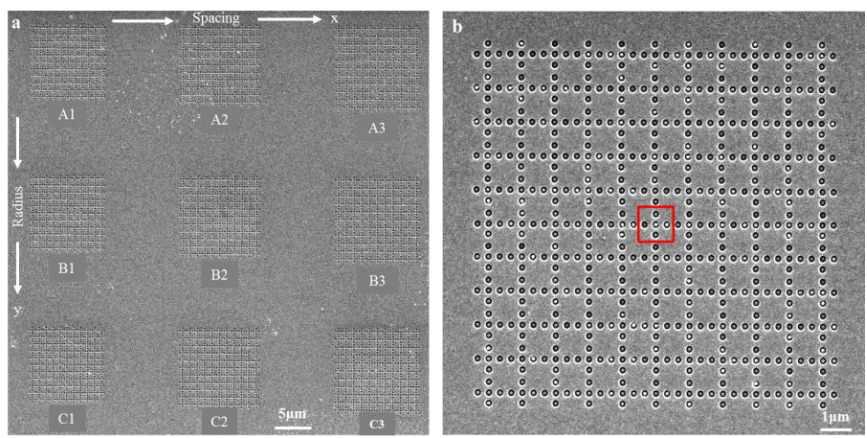

**Figure 3.** SEM image of the combinatorial meta-lens sample: (a) Matrix of nine samples with varying inter-ring spacing and ring radii. (b) Sample A3 consisting of 11x11 meta-lens units; a single unit is marked with red box.

The sample used for demonstrating the proof-of-principle superoscillatory meta-lens consisted of a combination of nine different meta-lens arrays, each with 11x11 meta-lenses, with varying parameters. In figure 3a, the radii of constituent ring slots measures 75 nm for all meta-lens arrays along the top row and increases in 5 nm steps for each successive row. The centre-to-centre distance between the neighbouring meta-atoms is 300 nm for all meta-lens arrays on the left most column and increases in 20 nm steps for consecutive columns. The meta-lens arrays were fabricated by focused ion beam milling of a 50 nm thin gold film deposited on a glass substrate by resistive evaporation. For all the rings the groves were 34 nm wide.

The performance of the meta-lenses was investigated for 750, 800 and 850 nm wavelengths with a circularly polarized laser illuminating the arrays from the substrate side. The intensity on the other side of the arrays was imaged in immersion oil by a CDD camera through a liquid immersion microscope objective (NA=1.4). The formation of foci and the reconstruction of the array at periodic distances from the meta-lens arrays were observed. We present here the best result corresponding to 800 nm illumination on meta-lens structure A3 shown in figure 3b.

Figure 4 presents a summary of the experimental characterization of meta-lens A3. The intensity distribution on the surface of the sample, when illuminated with plane wave, is shown on figure 4a. The foci above each meta-lens unit were observed to re-appear after every 2.8 μm along the propagation direction. This is presented on figure 4b where the intensity distribution directly above a meta-lens unit is plotted along the propagation direction. The repetition distance between consecutive focal spots (Talbot distance) as observed from the simulation differs from the experimental measurement by about 0.4 μm. This slight discrepancy may be attributed to unintentional spherical wavefront illumination in the experimental set-up which changes Talbot distances [14] compared to plane wave illumination.

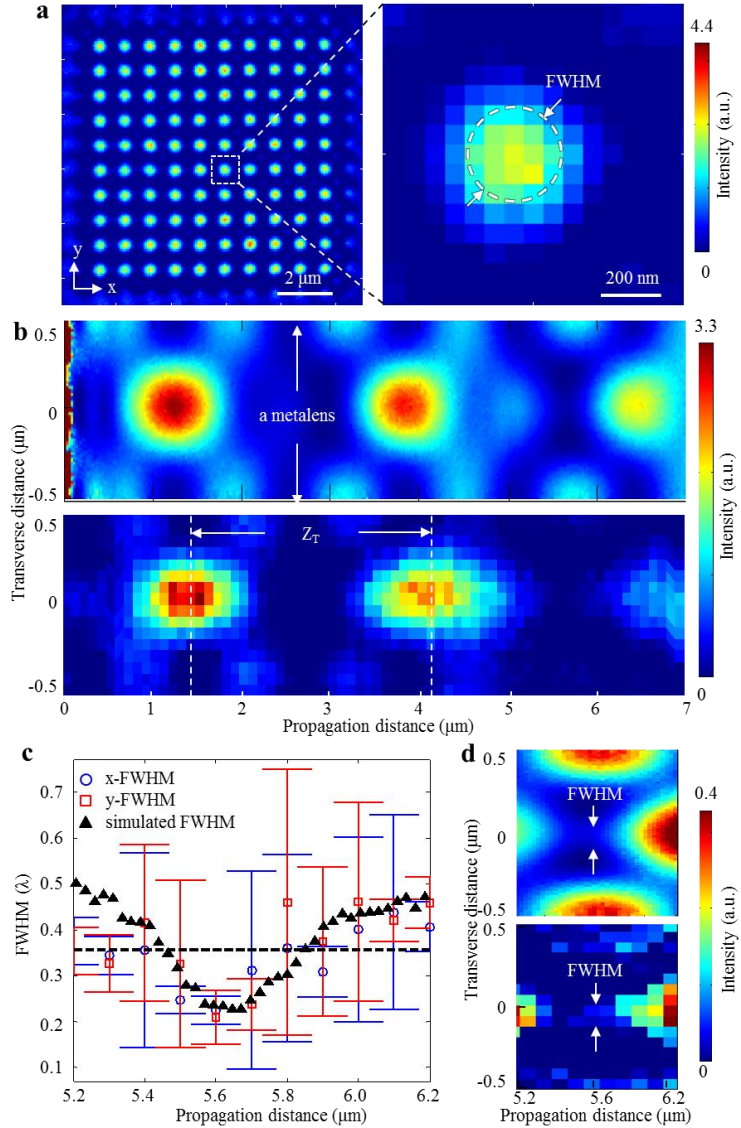

**Figure 4.** Experimental focussing with meta-lens A3: (a) Intensity distribution on the surface of the meta-lens array; (right) over a meta-lens unit. (b) Simulated (top) and experimental (bottom) intensity plots over one meta-lens unit showing repeated formation of focal spots along the propagation direction with Talbot distance $z_T$. (c) FWHM of a spot vs. the propagation distance. The focal spot is below diffraction limit (dashed line). (d) Zoom of (b) showing low intensity region over which focal spot is smaller than the diffraction limit.

To analyze the sub-wavelength characteristics of the hot-spots, a single spot directly over a meta-lens unit was chosen and the full width at half maximum (FWHM) was measured for each position along the propagation direction. Figure 4c shows a plot of the FWHM, as measured and simulated, along the propagation direction where small spot sizes were observed. The error bars give standard deviation of FWHM measured for five different spots on each plane parallel to the sample surface. The smallest spot (FWHM = 176 nm) occurred at 5.6 μm with low intensity levels (figure 4d). In fact, all the spots smaller than the diffraction limit (dotted line in figure 4c) were found in the low intensity region and several wavelengths away from the meta-lens surface which are characteristic of optical superoscillations [2].

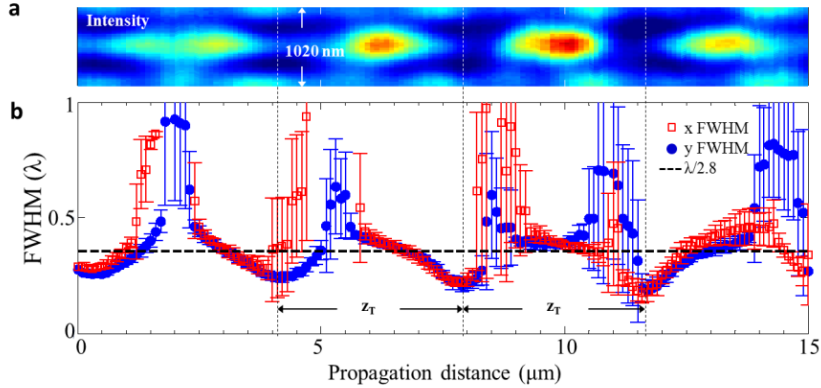

**Figure 5.** Long-range performance of a superoscillatory meta-device: (a) Experimentally measured intensity and (b) FWHM from a sample formed by 30 by 30 meta-lenses. Hot-spots measuring 0.2λ at 11.7 μm from the lens with Talbot distance of 3.8 μm are observed.

In the experiment, due to the finite size of each of the samples in the combinatorial array, the distance up to which the hot-spots could be characterized was limited by strong edge diffraction effects. To address this, a large array consisting of 30 by 30 meta-lenses was fabricated with each meta-lens unit measuring 1.02 μm by 1.02 μm. When illuminated with an 800 nm circularly polarized laser, hot-spots measuring as small as 160 nm was observed as far as 11.7 μm from the metamaterial lens surface (figure 5b). The hot-spots were repeated along the propagation distance with a Talbot distance of 3.8 μm. As in the case of the smaller array of meta-lenses, the smallest spots always appeared in the low intensity region; the size of the high intensity spots is around the diffraction limit.

In conclusion, we report a superoscillatory metamaterial lens made of clusters of ring grooves on a thin metal film that can act as a focusing device depending on the parameters of the meta-atoms and optical wavelength. We have also identified the condition when the focusing device can generate arrays of sub-wavelength foci as small as λ/5 at a distance 14.7λ from the meta-lens surface. Such planar arrays of meta-lenses could find applications in high resolution parallel imaging, photolithography, and data storage.